**Tinetti G., et al. 2007, Nature, 448, 163.**

# Water vapour in the atmosphere of a transiting extrasolar planet

Giovanna Tinetti[1,2,3], Alfred Vidal-Madjar[3], Mao-Chang Liang[4,5], Jean-Philippe Beaulieu[3], Yuk Yung[5], Sean Carey[6], Robert J. Barber[2], Jonathan Tennyson[2], Ignasi Ribas[7], Nicole Allard[3], Gilda E. Ballester[8], David K. Sing[3,9], & Franck Selsis[10]

[1] *European Space Agency*

[2] *Department of Physics and Astronomy, University College London, Gower Street, London WC1E 6BT, UK*

[3] *Institut d'Astrophysique de Paris, CNRS, Université Pierre et Marie Curie, 75014 Paris, France.*

[4] *Research Center for Environmental Changes, Academia Sinica, Taipei 115, Taiwan*

[5] *California Institute of Technology, Division of Geological and Planetary Sciences, Pasadena, CA 91125, USA*

[6] *California Institute of Technology, IPAC-Spitzer Science Center, Pasadena, CA 91125, USA*

[7] *Institut de Ciències de l'Espai (CSIC-IEEC), Campus UAB, 08193 Bellaterra, Spain*

[8] *Department of Planetary Sciences, Lunar and Planetary Laboratory, University of Arizona, USA*

[9] *Centre National d'Etudes Spatiales, France*

[10] *Ecole Normale Supérieure, CNRS, Lyon, France*

**Water is predicted to be among, if not the most abundant molecular species after hydrogen in the atmospheres of close-in extrasolar giant planets (hot-Jupiters)[1,2]. Several attempts have been made to detect water on an exoplanet, but have failed to find compelling evidence for it[3,4] or led to claims that should be taken with caution[5]. Here we report an analysis of recent observations of the hot-Jupiter HD189733b[6] taken during the transit, where the planet passed in front of its parent star. We find that absorption by water vapour is the most likely cause of the wavelength-dependent variations in the effective radius of the planet at the infrared wavelengths 3.6, 5.8 (ref. 7) and 8 μm (ref. 8). The larger effective radius observed at visible wavelengths[9] may be due to either star variability or the presence of clouds/hazes. We explain the most recent thermal infrared observations of the planet during secondary transit behind the star, reporting a non-detection of water on HD189733b[4], as being a consequence of the nearly isothermal vertical profile of the planet's atmosphere. Our results show that water is detectable on extrasolar planets using the primary transit technique and that the infrared should be a better wavelength region than the visible, for such searches.**

Water absorbs over a broad wavelength range, covering most of the infrared and part of the visible, and has a very distinctive spectral signature. Theoretical models anticipated that it could be detected on hot Jupiters by observing these planets during their primary transit (when they pass in front of their parent star) and estimating the effective radius of the planet at multiple wavelengths. Variations in this effective radius can be used to characterise the transmission spectrum of the planet's atmosphere[1,10,11]. This technique had been successfully used to probe other atmospheric constituents on hot-Jupiters[12-15]. Recently water detection was claimed[5] in the atmosphere of the hot-Jupiter HD209458b. Although it is possible that the observed variations in the effective radius of the planet in the visible/near-infrared (~0.9-1 μm)[16] are due to the absorption of water[10], the analysis is based on the very lowest-flux part of the spectrum, at the edge of the detector array where the largest systematic effects occur (Fig. 3 in ref. 16) and the overall noise is much larger. The quoted errors in the effective radius are based on photon noise alone, and do not include these systematic uncertainties[16]. These considerations alone are a good reason for caution.

Very recently the planet to star radius ratios of another hot-Jupiter, HD189733b[6], were measured in three of the Spitzer Space Telescope-InfraRed Array Camera (IRAC) bands centred at 3.6 (ref. 7), 5.8 (ref. 7) and 8 μm (ref. 8). The corresponding transit depths, i.e. the ratio of the projected area of the planet to that of the star, were estimated to be 2.356±0.02% (ref. 7), 2.436±0.02% (ref. 7) and 2.39±0.02% (ref. 8). To interpret these data, we simulated transmission spectra, which determine the effective radius of the planet at different wavelengths, by improving upon a previous spectral/planetary model[11]. The radiative transfer and geometry are unchanged from that model: we use 45 atmospheric layers to describe the variation with altitude of the temperature, pressure (from 10 bar to ~$10^{-10}$ bar, corresponding to ~4-6·$10^3$ km altitude depending on the thermal profile), density and mixing ratios. The water volume mixing ratio is about 5·$10^{-4}$, assuming an elementary C/O ratio equal to solar abundance[2]. Sensitivity studies to the changes of water abundances are presented. The crucial difference with the previous model is the use of a new high-accuracy computed water line list, BT2[17]. While only about 80,000 water line strengths are known experimentally, the BT2 line list contains more than half a billion transitions; these extra transitions become increasingly important at higher temperatures. As a result, the BT2 opacity at a particular wavelength is far more temperature-dependent than might be expected from computations employing only experimental data. The opacities were calculated for the selected spectral band at different temperatures from 500 K to 2000 K, and interpolated for intermediate values of the temperature at each atmospheric layer.

We repeated our simulations with different thermal profiles, compatible with 3D climate models of hot-Jupiters for the day/night sides and morning/evening terminators[18]. These simulations predict adiabatic profiles colder at the µbar level (~500-700 K) and warmer at 1-10 bars (~1500 K) for the night/terminator and nearly isothermal profiles in the upper (~800-1000 K) and lower part of the atmosphere (~2000 K) for the day side. The temperature might increase again (~2000 K) at pressures ~$10^{-6}$-$10^{-10}$ bar. This increase is consistent with both models and observations of the planets in our solar system and beyond[15]. We used photochemical models[2,19] with updated nitrogen chemistry from NIST database (http://www.nist.gov) to estimate the ammonia abundance. In general, the predicted ammonia mixing ratios are $\ll 10^{-7}$. The ammonia absorption coefficients were estimated using HITRAN[20] data corrected for the higher temperatures. This additional absorption by ammonia makes negligible contribution to the observed infrared absorptions. CO and $CH_4$ also absorb in the spectral range considered. CO if present in sufficient abundance to be detected[2], should show its signature in the 4.5µm IRAC band[11], not yet observed in primary transit. $CH_4$ is ruled out both by the relative contribution of the observed IRAC bands and by the predictions of the photochemical models[2].

In our simulations, sodium and potassium were included with solar abundances. The line shapes of these alkali metals were calculated at different temperatures and interpolated for intermediate values[21,22]. Their spectral contribution becomes important in the visible wavelength range (Fig. 2). The $H_2$-$H_2$ opacity[23] was interpolated to the temperature of each atmospheric layer. As collision induced absorption scales with the square of the pressure, the $H_2$-$H_2$ contribution becomes important for pressures higher than ~ 1 bar.

The difference in absorption depths found in the IRAC bands centred at 3.6, 5.8 and 8 µm can be explained by the presence of water vapour in the atmosphere of HD189733b. Fig. 1 compares our calculated water absorption with the observations. The sensitivity of our simulations to temperature and water abundance is also considered in the figure. The scenarios which match the observations better, are terminator profiles[18]. These profiles (see bold curves in Fig. 4 in ref. 18) are in agreement with recent estimates of the hemisphere-averaged brightness temperature[8].

We extended our simulations of the transmission spectra from the infrared to the visible to check the consistency of the current interpretation with the observations at shorter wavelengths[9]. The measure of the transit depth in the optical was found to be 2.48±0.05% (ref. 9), higher than the infrared values. The extra absorption in the visible is probably due to the effect of star spots[9] and/or by the presence of optically thick clouds/hazes in the visible wavelength range (see Fig. 2). Large stellar spots or a condensate/haze where the particle-size is less than 1 μm, may not affect the infrared transmission spectrum[10,24]. This scenario is in agreement with models and observations of the planet HD209458b, which has characteristics similar to planet HD189733b considered here[10]. For HD209458b, the transit depth at 24 μm (ref. 25) does not differ significantly from the visible results. This observational constraint is consistent with our simulations for HD189733b (Fig. 2). Our explanation differs from a recent paper which proposes a cloud-free atmosphere with rainout and photo-ionization to fit multiband photometry measurements of HD209458b in the optical[5].

Finally, we simulated the emission spectra of HD189733b with the same atmospheric constituents but different thermal profiles. These synthetic spectra are useful for comparing our analysis of the primary transit observations, which indicate the presence of water vapour in the atmosphere of HD189733b, with the most recent secondary transit observations of HD189733b and HD209458b made with the Spitzer-Infrared Spectrograph (IRS)[3,4] which failed to detect water in the infrared emission spectra in the 7.5-14 μm region. Secondary transit is another technique to study the characteristics of hot-Jupiters' atmospheres. This method involves collecting the photons emitted directly by the planet and measuring the brightness of the star plus planet system and its progressive dimming when the planet is occulted by the parent star[26,27]. The previous non-detection of water on HD189733b[4] does not contradict the interpretation we present here. Secondary transit observations require a significant temperature gradient to be sensitive to atmospheric molecules absorbing in the infrared, but strong circulation on hot-Jupiters can flatten the day side temperature gradient[28]. This phenomenon is illustrated in a classic paper[29] that gives the first portrait of the Earth in the thermal infrared from space: despite the fact that $CO_2$ is nearly uniform in the atmosphere, the 667 $cm^{-1}$ $CO_2$ feature is prominent in the tropics, where the thermal gradient between the surface and the upper atmosphere is large (~100 K), but it is hardly detectable in polar observations, where stratospheric and surface temperatures are nearly the same[24]. A simple model of HD189733b infrared emission spectra illustrating this effect is given in Fig. 3. We note a recent different explanation[30]: the inconsistency of the spectral data for HD189733b obtained with IRS[4] with the 8 µm-photometry obtained with IRAC[8] may indicate that the IRS spectra from 7.5-10 µm are not completely reliable.

Our result provides the first detection of a molecular species in an exoplanet atmosphere in the infrared. Although the detection of water is secure, more observations at multiple wavelengths are needed to constrain its abundance. Moreover the IRAC band centred at 4.5 µm, could constrain the presence of CO on HD189733b, and indirectly the C/O ratio[11]. Additional observations in secondary transit in the infrared will help to refine the thermal profiles, the presence of condensates/hazes and their variability with time on the day side. Analogue observations are desirable for other hot-Jupiters to start comparative planetology for this class of objects.

1. Seager, S. and Sasselov, D.D., Theoretical Transmission Spectra during Extrasolar Giant Planet Transits. *Astrophys. J.* **537**, L916-L921 (2000).

2. Liang, M. C., Parkinson, C. D., Lee, A. Y.-T., Yung, Y.L., and Seager S. Source of Atomic Hydrogen in the atmosphere of HD209458b. *Astrophys. J.* **596**, L247-L250 (2003).

3. Richardson, L. J., Deming, D., Horning, K., Seager, S., Harrington, J., A spectrum of an Extrasolar planet. *Nature,* **445**, 892-895 (2007)

4. Grillmair, C.J., *et al.*, A Spitzer Spectrum of the Exoplanet HD189733b. *Astrophys. J.,* **658**, L 115-L118 (2007).

5. Barman, T., Identification of absorption features in an extrasolar planet atmosphere, *Astrophys. J.* **661,** L191-L194 (2007).

6. Bouchy, F., *et al.*, ELODIE metallicity-biased search for transiting Hot Jupiters II. A very hot Jupiter transiting the bright K star HD189733. *Astronomy and Astrophysics* **444**, L15-L19 (2005).

7. Beaulieu, J. P. *et al.*, Spitzer observations of the primary transit of the planet HD189733b at 3.6 and 5.8 μm. *Astrophys. J.* submitted.

8. Knutson, H. A. *et al.*, A map of the day-night contrast of the extrasolar planet HD189733b, *Nature,* **447**, 183-186 (2007).

9. Winn, J. N., *et al.*, The Transit Light Curve Project. V. System Parameters and Stellar Rotation Period of HD 189733. *Astron. J.* **133**, 1828-1835 (2007).

10. Brown, T. M., Transmission Spectra as Diagnostics of Extrasolar Giant Planet Atmospheres, *Astrophys. J.* **553**, 1006-1026 (2001).

11. Tinetti G., *et al.*, Infrared Transmission Spectra for Extrasolar Giant Planets, *Astrophys. J.* **654**, L99-L102 (2007).

12. Charbonneau, D., Brown, T.M., Noyes, R.W. & Gilliland, R.L. Detection of an Extrasolar Planet Atmosphere. *Astrophys. J.* **568**, 377-384 (2002).

13. Vidal-Madjar, A. *et al.* An Extended upper atmosphere around the extrasolar planet HD209458b, *Nature* **422**, 143-146 (2003).

14. Vidal-Madjar, A. *et al.* Detection of oxygen and carbon in the upper atmosphere of the extrasolar planet HD209458b, *Astrophys. J.* **604**, L69-L72 (2004).

15. Ballester, G. E., Sing, D. K., & Herbert, F., The signature of hot hydrogen in the atmosphere of the extrasolar planet HD 209458b *Nature*, **445**, 511-514 (2007).

16. Knutson, H. A., Charbonneau, D., Noyes, R.W., Brown, T. M., and Gilliland., R.L., Using Stellar Limb-Darkening to Refine the Properties of HD 209458b. *Astrophys. J.* **655,** 564–575 (2007)

17. Barber, R.J., Tennyson, J., Harris, G.J., Tolchenov, R., A high accuracy computed water line list. *Mon. Not. R. astr. Soc.* **368**, 1087-1094 (2006).

18. Burrows, A., Sudarsky, D., and Hubeny, I., Theory for the secondary eclipse fluxes, spectra, atmospheres and light curves of transiting extrasolar giant planets. *Astrophys. J.* **650**, 1140–1149 (2006).

19. Liang, M. C., Yung, Y. L., and Shemansky, D. E., Photolytically generated aerosols in the mesosphere and thermosphere of Titan. *Astrophys. J.,* **661** L199-L201 (2007).

20. Rothman, L. S. *et al.*, The HITRAN 2004 molecular spectroscopic database. *Journal of Quantitative Spectroscopy and Radiative Transfer* **96**, 139-204 (2005)

21. Allard, N. F., Allard, F., Hauschildt, P. H., Kielkopf, J. F., and Machin, L., A new model for brown dwarf spectra including accurate unified line shape theory for the Na I and K I resonance line profiles, *Astronomy and Astrophysics* **411**, L473–L476 (2003)

22. Allard, N. F., Spiegelman, F., and Kielkopf, J. F., Study of the K-$H_2$ quasi molecular line satellite in the potassium resonance line. *Astronomy and Astrophysics*, **465**, 1085-1091 (2007).

23. Borysow, A., Jorgensen, U. G., and Fu, Y., High temperature (1000-7000K) collision induced absorption of $H_2$ pairs computed from the first principles, with application to cool and dense stellar atmospheres, *Journal of Quantitative Spectroscopy and Radiative Transfer* **68**, 235-255 (2001).

24. Goody R.M., and Yung, Y.L., *Atmospheric Radiation*. Oxford University Press (1989).

25. Richardson, L. J., Harrington, J., Seager, S. and Deming, D., A Spitzer Infrared Radius for the Transiting Extrasolar Planet HD209458b. *Astrophys. J.* **649**, 1043-1047 (2006).

26. Charbonneau, D. *et al.* Detection of Thermal Emission from an Extrasolar Planet. *Astrophys. J.* **626**, 523-529 (2005).

27. Deming, D., Seager, S., Richardson, L. J, and Harrington J., Infrared radiation from an extrasolar planet. *Nature* **434**, 740 (2005).

28. Fortney, J.J., Cooper, C.S., Showman, A.P., Marley, M.S., and Freedman, R.S., The influence of atmospheric dynamics on the infrared spectra and light curves of hot Jupiters. *Astrophys. J.* **652**, 746-757 (2006)

29. Hanel, R. A., *et al.*, The Nimbus 4 Infrared Spectroscopy Experiment 1. Calibrated Thermal Emission Spectra. *J. Geophys. Res.* **77**, 2629- 2639 (1972).

30. Fortney, J. J. and Marley, M. S., Analysis of Spitzer Mid Infrared Spectra of Irradiated Planets: Evidence for Water Vapor? *Astrophys. J.,* submitted.

**Acknowledgements** A special thank to A. Noriega-Crespo and the Spitzer Staff to have helped scheduling the observations with IRAC, A. Lecavelier des Etangs, G. Hebrard, D. Ehrenreich, J.M. Desert, for their work on IRAC observations, E. Lellouch, A. Morbidelli, B. Schultz, F. Bouchy, J.B. Marquette for useful input to the paper. MCL and YLY were supported by NASA grant NASA5-13296 and the Virtual Planetary Laboratory at the California Institute of Technology.

Correspondence and requests for materials should be addressed to G.T. (giovanna@apl.ucl.ac.uk, gio@gps.caltech.edu).

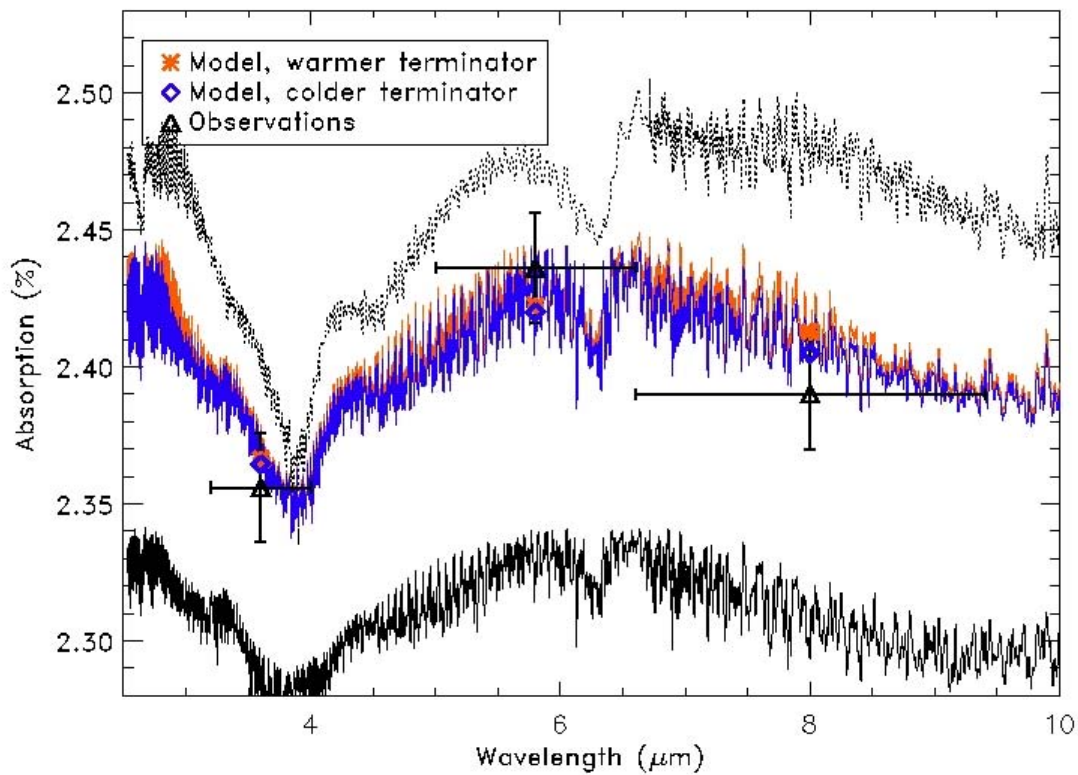

**Fig. 1: A comparison of the observations with simulated water absorption.** In these simulated transmission spectra, the water mixing ratio profile is assumed to be constant[2] and $\sim 5 \cdot 10^{-4}$. The observations are indicated with triangles and error bars at 1 σ, the coloured rhombi and stars indicate the different models integrated over the IRAC bands. To match the observations, the planetary radius at 10 bar corresponds to a transit depth of 2.28%. Blue plot: colder terminator[18] Temperature-Pressure (T-P) profile, orange plot: warmer terminator[18], black plots: constant temperature at 500 K (solid line) and 2000 K (dotted line) respectively. Warmer temperatures increase the atmospheric scale-height (i.e. the vertical distance over which the pressure decreases by a factor of *e*), hence the atmosphere is optically thick at higher altitudes. This explains the differences among the 3 classes of spectra at wavelengths shorter than ~3.5 μm and longer than ~4.5 μm, where the water opacities are far less temperature-dependent. The opposite is true for the water opacities in the 3.5-4.5 μm wavelength range, which might be order of magnitudes smaller at 500 K rather than at 2000 K, so for colder temperature profiles the weaker water lines are optically thick at ~10 bar or deeper. An increase/decrease of the mixing ratio of a factor 10 with respect to the standard case considered will cause, as a main effect, an increase/decrease of ~0.03-0.04% in the total absorption due to water. As a secondary effect, the absorption gradient between 3.6 and 5.8 μm gets steeper for lower water mixing ratios, but this trend is marginal compared to the role played by temperature. CO, if present in sufficient abundance to be detected[2], would show its distinctive signature in the 4.5-4.9 μm spectral range (see ref. 11 for details). This is a spectral region than can be observed with IRAC (channel centred at 4.5 μm).

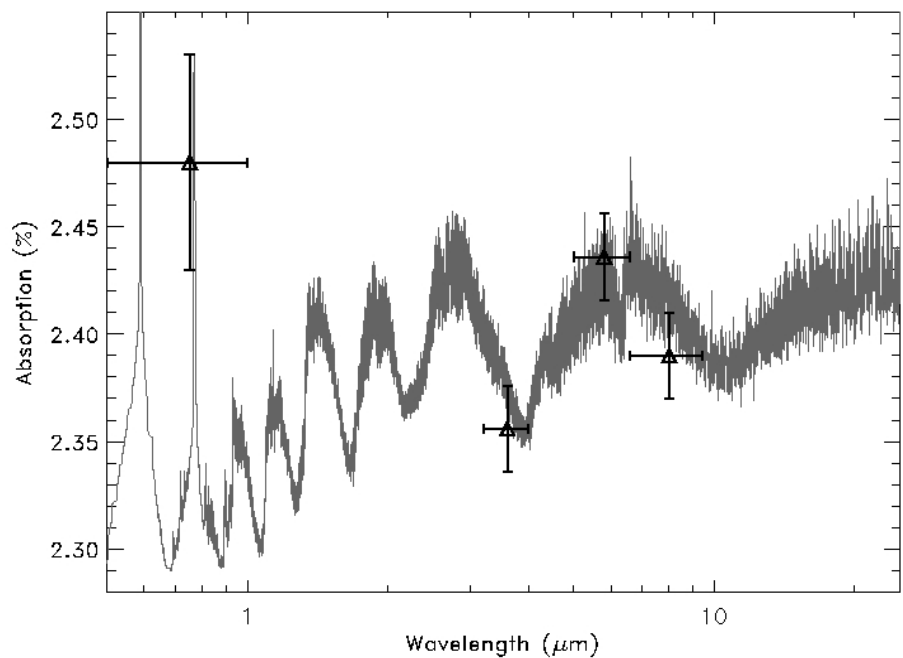

**Fig.2: A comparison of the observations of primary transit with a simulated infrared and optical transmission spectrum.** The transmission spectrum is here modelled from 0.5 to 25 μm (grey plot). We assumed a water mixing ratio of ~$5·10^{-4}$ and a cold terminator T-P profile. The observations are indicated with triangles and error bars at 1 σ. Alkali metals and water absorption are unlikely to be the cause of the extra-absorption observed at visible wavelengths: the mean absorption over the 0.5-1 μm band is significantly less than the measured value. Star spots or optically thick condensates/hazes would be a good explanation.

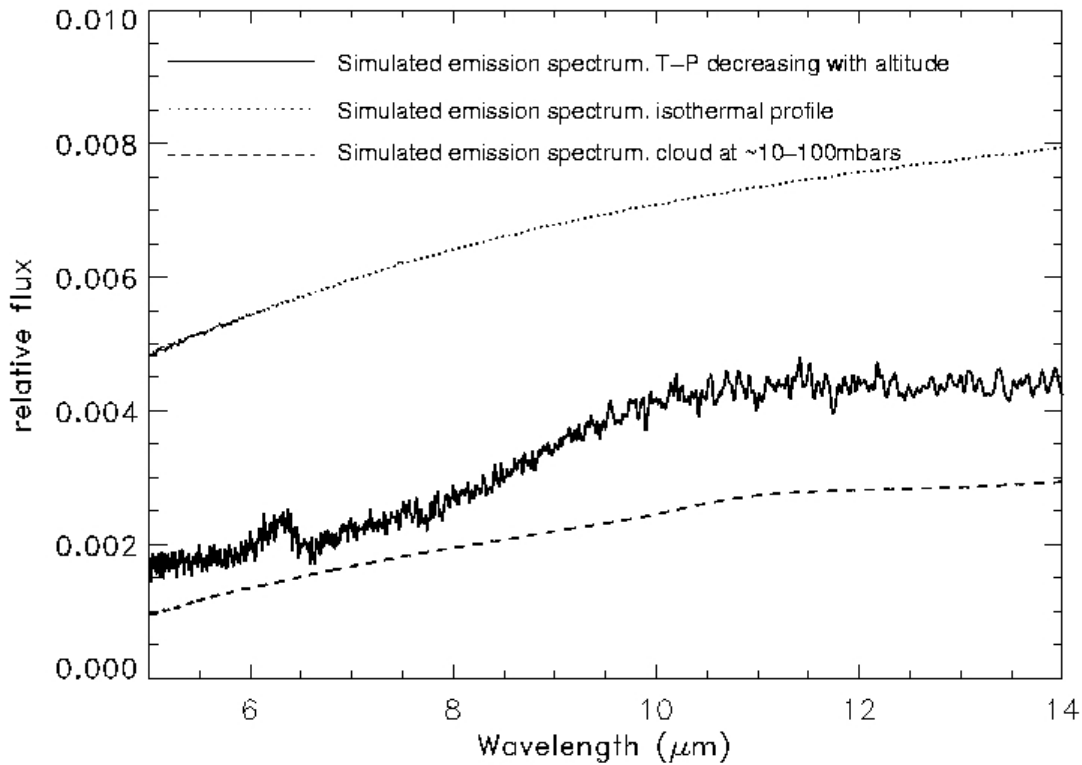

**Fig. 3: Simulated emission spectra of HD189733b in the infrared.** We show here different simulated emission spectra, all generated at the same spectral resolution and corresponding to the same atmospheric composition but different thermal profiles. Accordingly, we obtain very different spectral responses: hot isothermal profile (top), isothermal profile and a cloud at ~10-100 mbar altitude (bottom), temperature profile decreasing with altitude (middle). Although present and with the same abundances in all the three scenarios, water is detectable through emission spectra only in the last case, showing up a steep gradient between ~8-10 μm when compared to the 10-14 μm region. An isothermal profile in the upper part[18,28] and a cloud in the lower part of the atmosphere is sufficient to explain the non-detection of water[4] even if water were present in high abundance. HD189733b is presumably tidally locked, therefore the thermal profiles and the condensate dynamics might be very different on the two sides

of the planet. The same reasoning applies to HD209458b and the most recent observations in secondary transit of that planet[3].